\documentclass{ws-procs9x6}

\def\beq{\begin{equation}}
\def\eeq{\end{equation}}
\def\bea{\begin{eqnarray}}
\def\eea{\end{eqnarray}}

\def\dofigs#1#2#3{\centerline{\epsfxsize=#1\epsfbox{#2}%
   \hfil\epsfxsize=#1\epsfbox{#3}}}

\def\bsg{\ifmmode B_d\to X_s\gamma\else $B_d\to X_s\gamma$\fi}
\def\bsglue{\ifmmode B_d\to X_s\, g\else $B_d\to X_s\, g$\fi}
\def\bsll{\ifmmode B_d\to X_s\ell^+\ell^-\else $B_d\to X_s\ell^+\ell^-$\fi}
\def\bstt{\ifmmode B\to X_s\tau^+\tau^-\else $B\to X_s\tau^+\tau^-$\fi}
\def\bphik{\ifmmode B_d\to \phi K_s\else $B_d\to \phi K_s$\fi}
\def\bbarphik{\ifmmode \bar{B_d}\to \phi K_s\else $\bar{B_d}\to \phi K_s$\fi}
\def\bsmix{\ifmmode B_s \bar{B}_s\else $B_s \bar{B}_s$\fi}
\def\bdmix{\ifmmode B_d \bar{B}_d\else $B_d \bar{B}_d$\fi}
\def\bclnu{\ifmmode B_d\to X_c e \bar{\nu}\else $B\to X_c e \bar{\nu}$\fi}
\def\bjpsik{\ifmmode B_d\to J/\psi~K_s\else $B_d\to J/\psi~K_s$\fi}
\def\bsee{\ifmmode B_d\to X_s e^+ e^-\else $B_d\to X_s e^+ e^-$\fi}
\def\bsmumu{\ifmmode B_d\to X_s\mu^+\mu^-\else $B_d\to X_s \mu^+\mu^-$\fi}

\def\cal{ }  

\begin{document}
\title{
B-physics Signature of a \\
Supersymmetric U(2) Flavor Model\footnote{\uppercase{T}alk presented by 
\uppercase{Shrihari Gopalakrishna} at {\it \uppercase{SUSY} 2003:
\uppercase{S}upersymmetry in the \uppercase{D}esert}\/, 
held at the \uppercase{U}niversity of \uppercase{A}rizona,
\uppercase{T}ucson, \uppercase{AZ}, \uppercase{J}une 5-10, 2003.
\uppercase{T}o appear in the \uppercase{P}roceedings.}}

\author{Shrihari Gopalakrishna and C.--P. Yuan}
\address{
Department of Physics and Astronomy\\
Michigan State University\\ 
East Lansing, MI 48824 USA\\ 
E-mail: shri@pa.msu.edu, yuan@pa.msu.edu}



\maketitle

\abstracts{
We discuss the B-physics signature of a supersymmetric U(2) flavor model in which the 
third generation scalars are relatively light (electroweak scale masses). 
We impose current experimental constraints on such a framework and obtain expectations 
for various B-physics processes. Here we present CP violation in \bsg\ and \bphik, and,
\bsmix\ mixing. We show that if realized in nature, 
such a framework can be discovered in current and upcoming experiments.}

\section{Introduction}
The Standard Model (SM) of high energy physics suffers from the gauge hierarchy problem 
and the flavor problem. Supersymmetry (SUSY) eliminates the gauge hierarchy problem, and
a (horizontal) flavor symmetry in generation space could explain the flavor problem. 
A SUSY theory with a flavor symmetry might relate the quark/lepton flavor structure with that
of the scalar quark/lepton sector. Such a theory would imply certain predictions for flavor changing 
neutral current (FCNC) processes that we wish to investigate in this work, along with
the constraints from experimental FCNC data. 

We do not assume an alignment of the quark/lepton
flavor structure with that of the scalar quark/lepton sector, leading to a non-minimal flavor violation 
(NMFV) scenario. We consider a spontaneously broken U(2) flavor 
symmetry~\cite{Pomarol:1995xc,Barbieri:1995uv} in the
framework of ``effective supersymmetry''~\cite{Cohen:1996vb}, in which the first two generation 
scalars are relatively heavy (a few TeV mass), thereby satisfying neutron electric dipole moment 
constraint, etc., while still allowing large CP violating phases in the scalar sector. We analyze 
the implications of such a framework to B-physics observables. We will present details in a
forthcoming paper~\cite{shricp}.

Consider that the first and second generation superfields ($\psi_a$, a=1,2) transform as a 
U(2) doublet while the third generation superfield ($\psi$) is a singlet~\cite{Barbieri:1995uv}. 
The most general U(2) symmetric superpotential can be written as
\bea
{\cal W} = \psi \alpha_1 H \psi + \frac{\phi^a}{M} \psi \alpha_2 H \psi_a 
+ \frac{\phi^{ab}}{M} \psi_a \alpha_3 H \psi_b 
+ \frac{\phi^a \phi^b}{M^2}\psi_a \alpha_4 H \psi_b \\ \nonumber
+ \frac{S^{ab}}{M} \psi_a \alpha_5 H \psi_b + \mu H_u H_d \ ,
\eea
where $M$ is the cutoff scale below which such an effective description is valid, 
the $\alpha_i$ are O(1) constants, $\phi^a$ is a U(2) doublet, $\phi^{ab}$ and $S^{ab}$ 
are second rank antisymmetric and symmetric U(2) tensors respectively. 
If U(2) is broken spontaneously by the Vacuum Expectation Values (VEV) 
\bea
\left< \phi^a \right> = \begin{pmatrix}0 \\ V \end{pmatrix};\ \ \  \left<\phi^{ab}\right> = v \epsilon^{ab};\ \ \ \left<S^{11,12,21}\right> = 0, \left<S^{22}\right> = V ,  \nonumber
\eea
with $\frac{V}{M} \equiv \epsilon \sim 0.02$ and $\frac{v}{M} \equiv \epsilon ' \sim 0.004$, and
if U(2) is broken below the SUSY breaking scale, the SUSY breaking masses would also have a structure 
dictated by U(2). The resulting quark and scalar down-type masses are
\begin{eqnarray}
{\cal M}_d =  v_d \begin{pmatrix} O & -\lambda_1\epsilon ' & O \\ \lambda_1\epsilon '  & \lambda_2\epsilon & \lambda_4\epsilon \\  O & \lambda_4'\epsilon & \lambda_3 \end{pmatrix} \ &,& \quad 
{\cal M}^2_{RL} = v_d \begin{pmatrix}O & -A_1 \epsilon ' & O \\ A_1 \epsilon '  & A_2\epsilon & A_4 \epsilon \\  O & A_4'\epsilon & A_3 \end{pmatrix} \ , \\
{\cal M}^2_{LL} = \begin{pmatrix}m_1^2 & 0 & 0 \\
	 0 & m_1^2+\epsilon^2 m_2^2 & \epsilon m_4^{2*} \\ 
	0 & \epsilon m_4^2 & m_3^2 \end{pmatrix}_{LL}&,& \quad 
{\cal M}^2_{RR} = \begin{pmatrix} m_1^2 & 0 & 0 \\
	 0 & m_1^2+\epsilon^2 m_2^2 & \epsilon m_4^{2*} \\ 
	0 & \epsilon m_4^2 & m_3^2\end{pmatrix}_{RR}, \nonumber
\label{MSQ.EQ}
\end{eqnarray}
where $v_d = \left< h_d \right>$ is the VEV of the Higgs field, the $\lambda_i$'s are O(1) 
coefficients, and, $m_i$ and $A_i$ (complex in general) are determined by the SUSY 
breaking mechanism. It has been shown~\cite{Barbieri:1995uv} that such a pattern of the quark mass 
matrix explains the quark masses and CKM elements.

For our study, we consider the following values for the various SUSY parameters: 
${m_{\tilde b_R,\tilde t_R}}=100$\,GeV, the other squark masses given by $m_0=1000$\,GeV, 
$A=1000$\,GeV, $\tan{\beta}=5$, $|\mu|=150$\,GeV, $M_2=250$\,GeV, $M_{\tilde g}=250$\,GeV 
and $m_{H^\pm}=250$\,GeV. ($m_0$ and $A$ denote generic SUSY breaking mass scales.)

Here, we consider processes that go through the $b\rightarrow s$ quark level transition, 
and in our framework the dominant SUSY contributions are due to 
$\delta_{32,23}^{RL,RR,LL} \equiv \frac{({\cal M}^2_{RL,RR,LL})_{32,23}}{m_0^2}$. 
For the chosen values of the parameters, we find  
$|\delta_{32,23}^{RL}| \sim \frac{v_d A\epsilon}{\tilde{m}_0^2}=6.8\times 10^{-4}$, and,
$|\delta_{32}^{LL,RR}| \sim \epsilon \frac{m_4^2}{m_0^2} = 0.02$.

\section{B-physics probes}
Given such an effective SUSY theory we estimate the sizes of various B-physics observables 
that we expect are modified from their SM predictions.
In addition to the SM contribution, we include the charged Higgs, chargino and gluino
contributions. 
We analyze the $\Delta B=1$ FCNC processes, \bsg, \bsglue, \bsll, \bphik; and the 
 $\Delta B=2$ processes \bsmix\ mixing and the dilepton asymmetry in $B_s$.
We find regions in U(2) SUSY parameter space that are consistent with current experimental
data and obtain expectations for measurements that are forthcoming. To illustrate the effects, we
present here expectations for CP asymmetries in \bsg\ and \bphik, and, \bsmix\ mixing. A more 
exhaustive analysis will be presented elsewhere~\cite{shricp}.

\begin{figure}
\dofigs{2.2in}{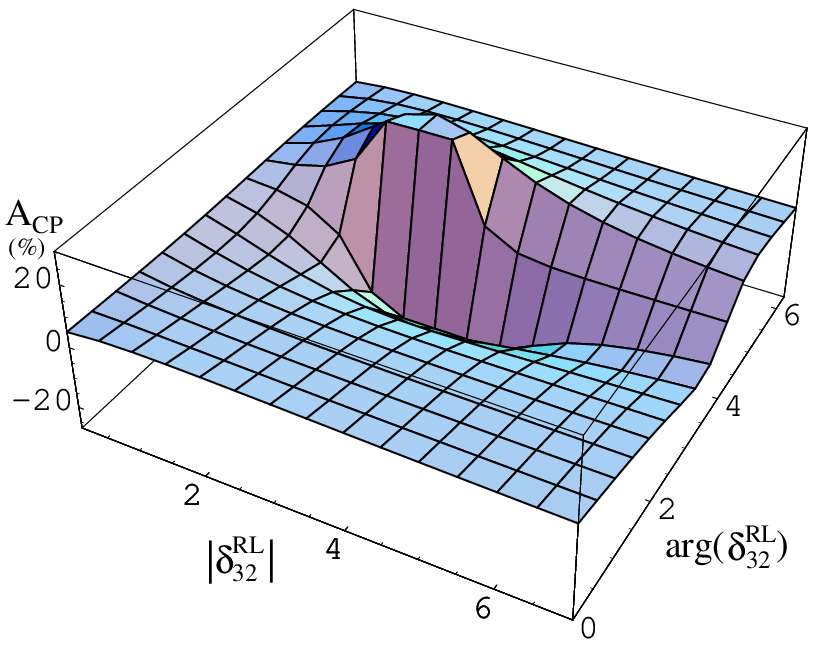}{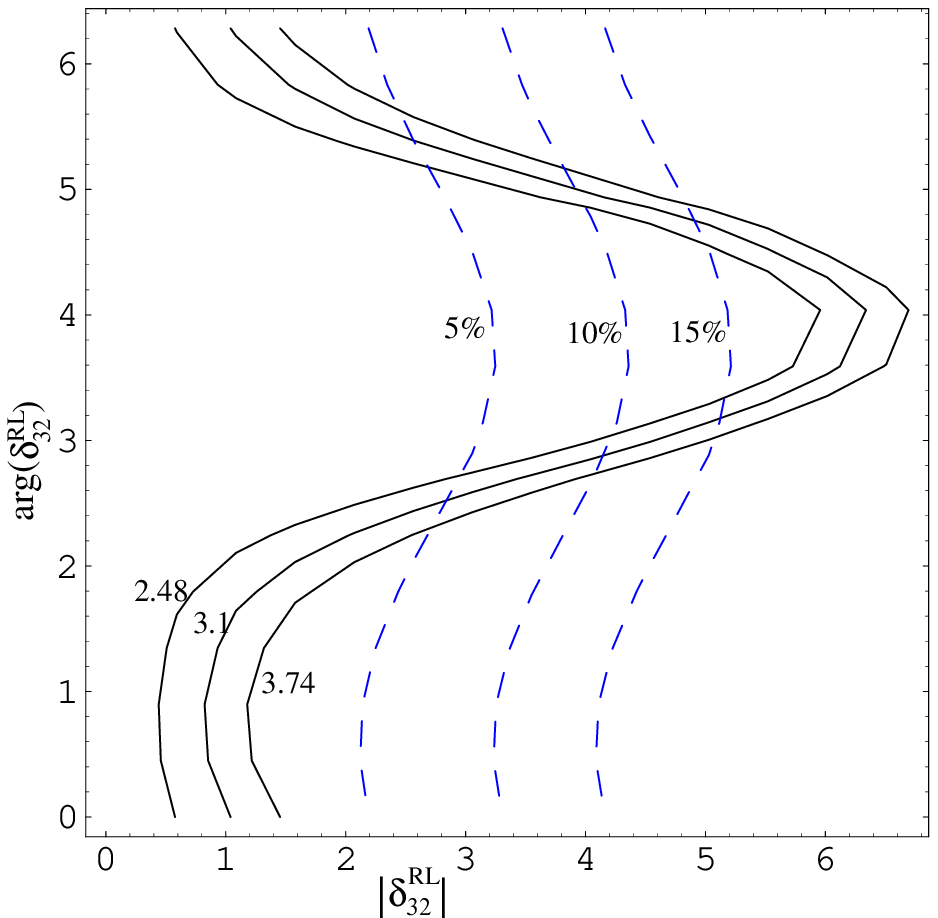}
\caption{
Left: $A_{CP}^{\bsg}$(in \%) as a function of $|\delta_{32}^{RL}|$(units of $6.8\times 10^{-4}$) and 
$\arg{(\delta_{32}^{RL})}$. 
Right: $1\,\sigma$ contours of B.R.(\bsg)~(solid lines, in units of $\times 10^{-4}$) and 5\,\%, 10\,\% and 15\,\% contours of B.R.(\bsglue)~(dashed lines). 
The other parameter values are as given in the text and $\arg{(\mu)}=5.4$.
}
\label{BSGACPLR.FIG}
\end{figure}

The CP asymmetry in \bsg\ is given by
\bea
A_{CP}^{\bsg} &\equiv&  \frac{\Gamma(\bar B_d \to X_s \gamma) - \Gamma(B_d \to X_{\bar s} \gamma)} {\Gamma(\bar B_d \to X_s \gamma) + \Gamma(B_d \to X_{\bar s} \gamma) } \ , 
\label{ACPBSG.EQ}
\eea
and the expectation in the U(2) SUSY theory is shown in Fig.~(\ref{BSGACPLR.FIG}). We see that 
significant CP asymmetry is possible in the scenario we are considering, while satisfying 
experimental constraints. 

\begin{figure}
\dofigs{2.2in}{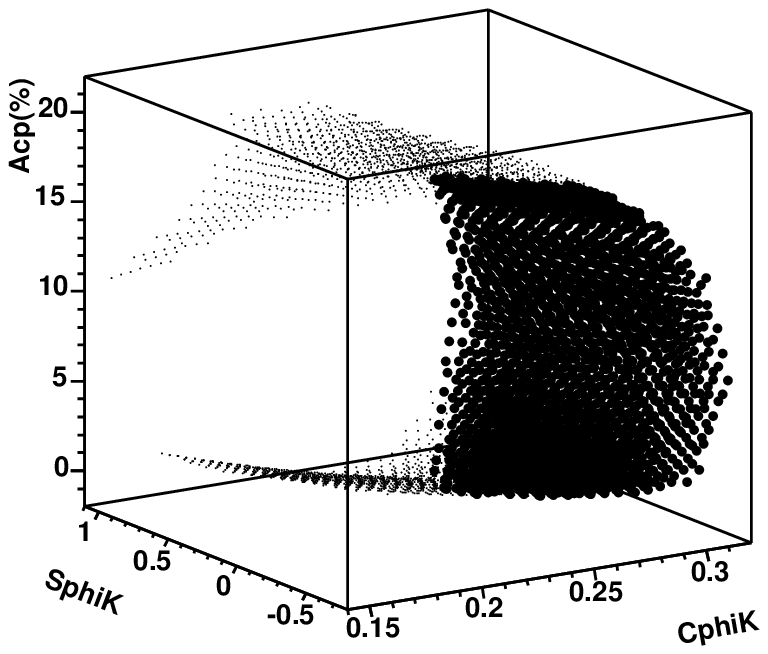}{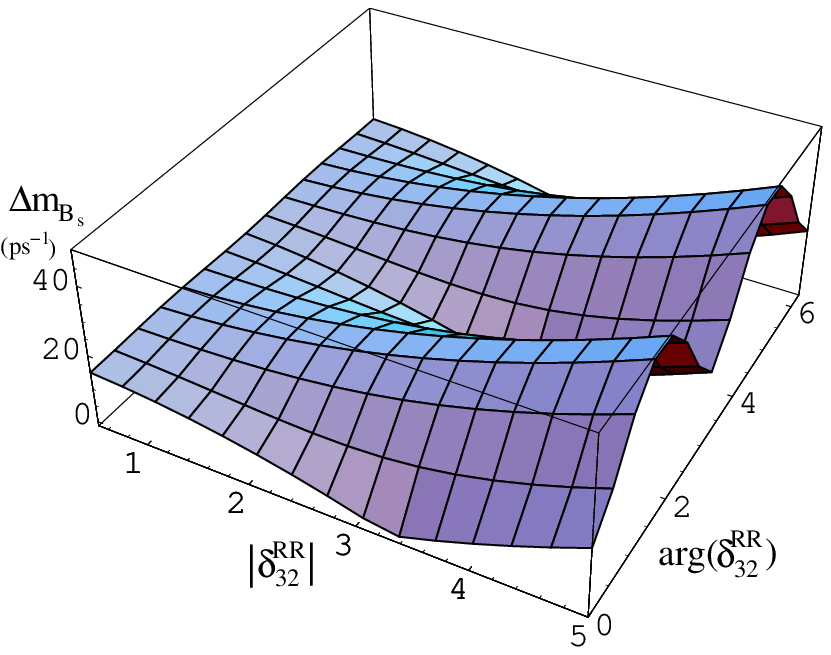}
\caption{Left: $A_{CP}^{\bsg}$, $S_{\phi K}$ and $C_{\phi K}$ for the region of parameter 
space that satisfy the experimental constraints of B.R.(\bsg) and B.R.(\bphik) 
(thin and thick dots) and those that also satisfy the current constraints on $S_{\phi K}$ and 
$C_{\phi K}$ (thick dots).  Right: The dependence of $\Delta m_{B_s}$ (ps$^{-1}$) on $|\delta_{32}^{RR}|$ 
(in units of $0.02$) and $\arg{(\delta_{32}^{RR})}$. 
}
\label{BPHIKCPBS.FIG}
\end{figure}

The CP asymmetry in \bphik\ is defined by
\bea
A_{CP}^{\bphik} &\equiv& \frac{\Gamma\left(\bar{B}_d(t)\to \phi K_s\right) - \Gamma\left(B_d(t)\to \phi K_s \right)}{\Gamma\left(\bar{B}_d(t)\to \phi K_s\right) + \Gamma\left(B_d(t)\to \phi K_s \right)} \\
&=& -C_{\phi K}\cos{\left(\Delta m_{B_d}\,t\right)} + S_{\phi K}\sin{\left(\Delta m_{B_d}\,t\right)},  
\eea
and Fig.~(\ref{BPHIKCPBS.FIG}~left) shows the CP asymmetry in \bphik\ for a scan on 
$\delta_{32}^{RL}$ while satisfying all experimental constraints.

The \bsmix\ mixing parameter $\Delta m_{B_s}$ depends quite sensitively on $\delta_{32}^{RR}$ 
and can be significantly altered from the SM prediction as shown in Fig.~(\ref{BPHIKCPBS.FIG}~right). 

In conclusion, we note that similar results hold for flavor models that have the same order 
of magnitude for the 23 element in the squark mass matrix. In such cases, the prospects 
look exciting for discovering SUSY in B meson processes at current and upcoming colliders. 


\end{document}